\documentstyle[aaspp4]{article}

\begin{document}
\title{Optical SETI with Air Cerenkov Telescopes}
\author{David Eichler\\Dept. of Physics\\Ben Gurion
University\\email: {eichler@bgumail.bgu.ac.il}}
\author{Gregory Beskin\\Special Astrophysical Observatory, Nizniji Arkhyz,
Russia
\\Isaac Newton Institute of Chile, SAO Branch
\\email: {beskin@sao.ru}}

\begin{abstract}
 We propose using large Air Cerenkov Telescopes (ACT's) to
search for optical, pulsed signals from extra-terrestrial
intelligence.  Such dishes collect tens of photons from a
nanosecond-scale pulse of isotropic equivalent power of tens of
solar luminosities at a distance of 100 pc.  The field of view for
giant ACT's can be on the order of ten square degrees, and they
will be able to monitor 10 to 10$^2$ stars simultaneously for
nanosecond pulses of about 6th mag or brighter.  Using the Earth's
diameter as a baseline, orbital motion of the planet could be
detected by timing the pulse arrival times.
\end{abstract}

\section{Introduction}
 Optical searches for extra-terrestrial intelligence
(ETI) were first proposed by Schwartz and Townes (1961) not long
after the invention of the laser.  They have been invigorated by
the realization that a civilization with Earth 2000 technology and
a sun similar to our own could, for a brief enough interval,
outshine its sun in the optical in a sufficiently well collimated
direction.  For example, a 30 kilojoule pulse, lasting 1
nanosecond with a $10^{-13}$ ster beam, gives an isotropic
equivalent power along the beam axis of 10 solar (bolometric)
luminosities. Thirty gigawatt-years of such pulses would fill a
fair fraction of the sky.  A civilization that expended an average
of 30 gigawatts (roughly 1/10 of the average U.S. electrical
consumption) in such pulses for, say, a star-wars type defense
system, might thus become inadvertently detectable.  Far less
energy would be required for us to be within such a beam if the
signals were deliberately directed at other stars like our Sun.
Thus, a civilization might choose to broadcast its existence with
brief optical pulses.  OSETI programs that use modest astronomical
telescopes with angular resolution of order 1 arcsec are underway
(Shvartsman $et\; al$. 1993, Beskin $et\; al$. 1997, Horowitz
$et\; al$. 2001). Below, we suggest that optical dishes with
poorer angular resolution may offer an alternative strategy for
OSETI.

The question then arises as to what sort of pulse would be
required to exclude the possibility of a natural source (which,
one might argue, a self-proclaiming civilization would
deliberately choose in order to avoid ambiguity).  At the very
least, they would be sure to make the pulses outshine their sun by
a comfortable margin that would allow unambiguous detection.  In
fact, existing OSETI teams have proposed searching for megajoule
pulses that could outshine our sun by a factor of 10$^3$ or more.
However, the strong constraints posed by ecological considerations
and the desire to broadcast as much information as possible might
cause them to opt for weaker pulses.  At a distance of 100 pc, a
solar luminosity, L$_{solar}$, produces a flux at Earth of about
$10^{-6}$ optical photons per cm$^2$ per nanosecond.  We assume,
then, that an ETI would produce the minimum signal that would
stand out comfortably above this flux.

In this paper we consider the possibility of monitoring many stars
at once in an OSETI program using air Cerenkov telescopes (ACT's).
These dishes, planned for ground based gamma ray astronomy, are to
be outfitted with nanosecond imaging cameras. They offer the
following advantages:
\newline
1) Greater sensitivity, because of their large size ($\sim$ 10
meter diameters)
\newline
2) Large field of view (up to $\sim$ 10 square degrees), which,
combined with the first advantage, would allow monitoring of many
stars at once for sufficiently strong pulses.
\newline
3) Since the dishes will be built for ground based gamma ray
astronomy, OSETI programs could be implemented at some level by
introducing appropriate data storage and processing.

A disadvantage, however, is that a significant amount of empty sky
would be monitored, and this would require additional effort in
noise rejection, as discussed below.

\section{Materials and Methods}

 Giant optical dishes such as the planned air
Cerenkov telescope (ACT) MAGIC (Major Atmospheric Gamma Imaging
Cerenkov telescope, 236 m$^2$) (Barrio $et\; al$. 1998) and the
recently constructed PETAL (Photon Energy Transformer and
Astrophysical Laboratory, 400 m$^2$) dish (see figure) are, as far
as we know, the largest optical reflecting dishes in the world.
Arrays of ACT's, such as VERITAS (Very Energetic Radiation Imaging
Telescope Array System) and HESS (High Energy Stereoscopic
System), will have total areas of up to $\sim$ 10$^3$ m$^{2}$,
distributed over  many (4  to 10 or so) dishes.  They are quite
inexpensive per unit collecting area in comparison to astronomical
telescopes.  The MAGIC surface, diamond-turned aluminum, will cost
roughly US$\$1K$ per square meter, and can provide angular
resolution of about 1.5 arcminutes.  The typical pixel size for
the camera, however, will be on the order of 0.1 degree in
diameter for the inner pixels and about 0.2 degree for the outer
ones.  The PETAL surface, backsilvered glass, has an angular
resolution on the order of 0.2 degree, and costs $\sim$US$\$100$
per square meter. The surface is globally paraboloidal, so that
nanosecond pulses are not smeared by significant spread in path
lengths to the focus. The most expensive item in the latter case
is the molds, and we believe that resurfacing the dish with
polished sheet metal shaped on the existing molds could improve
the angular resolution at a cost of tens of US dollars per square
meter.

We now consider the required thresholds for signals.  To record a
small number of actual nanosecond pulses per year ($\sim$ 3 x
10$^{16}$ nanoseconds), we assume that false signals must be
reduced to one part in 10$^{17}$ as a standard of confidence.  The
diffuse night sky background under good conditions, about 22.5 mag
star per square arc second, translates to about 1.3 x 10$^{-7}$
photons/cm$^2$ nanosecond in the blue (B) band in a typical pixel
size of 30 square arcminutes (somewhat greater than a solar
luminosity at 100 pc, about 10th magnitude, which would give about
0.08 photons in the blue band for a standard solar spectrum).  If
this were the only background, assuming Poisson statistics, a 100
m$^2$ collecting area would register a clear signal (less than one
false event per year) with 11 B band photons arriving in a
nanosecond bin.  This would correspond to about 140 L$_{solar}$ at
100 pc.  An area much lower than 100 m$^2$ would give insufficient
statistics for the large detection confidence that the problem
requires.  A weaker signal would require a smaller pixel size to
surmount the night sky background. However, the  smaller pixel
size would only help if the host star of the extra-terrestrial
civilization were fainter than 10th magnitude.  Similarly,
restricting the search to faint host stars would not help, given
the diffuse background, if the pixel size is kept at about 30
square arcminutes. Altogether, it seems   that typical parameters
for large ACT's  are well suited for OSETI out to 100 pc.

Using a value of 1.3 x 10$^{-7}$ photons/cm$^2$ nanosecond in the
blue band for the background photons rate per pixel, we have found
that a clear signal from a nanosecond laser pulse with isotropic
equivalent blue band luminosity of much less than 140 L$_{solar}$
at 100 pc would require many hundreds of square meters of net
collecting area (i.e. factoring in quantum efficiency).  We
estimate that the minimum isotropic equivalent luminosity that
could be detected above noise with ACT's of  three or four hundred
m$^2$ would correspond in the B band to about 50 to 70 L$_{solar}$
at 100 pc, or about a 6th magnitude star.  We find that for 1000
m$^2$, at least 26 L$_{solar}$ at 100 pc is required.  We also
find that the minimum detectable intensity is slightly less in the
red (R) band; though the signal to noise is less in the red band,
the higher photon count rate compensates for this.  The
(presumably) monochromatic spectrum  of the laser pulse would work
in favor of picking it out of the background,  and  the above
estimates, if expressed in solar bolometric luminosities, would be
about an order of magnitude lower, depending on the details of the
detection scheme.

Air Cerenkov flashes produced by cosmic rays and gamma rays
typically have durations of 3 to 5 nanoseconds and present a major
noise problem.  Although they typically produce an imageable track
that covers many pixels, low energy hadronic cosmic rays (protons
in the 10 to 100 GeV range) may produce individual pions that
might produce quasi-pointlike clumps of optical photons.  These
clumps can contain on the order of 10 photons and lie within a
single pixel.  Though they are generally accompanied by many other
photons in other pixels, it is possible that, at the energy at
which pion production is marginal, single clumps would
occasionally masquerade as pointlike optical flashes. The arrival
rate of 30 GeV protons in the ACT field of view (some 10$^{-3.5}$
of the sky with a collecting area of about 5 x 10$^8$ cm$^2$) is
about 6 x 10$^3$/second. Even if one in ten such protons produces
a pointlike optical flash that gets past image selection, this
amounts to a probability of nearly 10$^{-6}$ per nanosecond for a
chance event. Such noise events can be selected out by
stereoscopic selection by using three to four suitably spaced
dishes in coincidence.  Arrays of ACT's, such as VERITAS and HESS,
seem suitable for this technique if appropriate trigger criteria
and software are implemented.

Yet another noise source is cosmic rays passing through individual
pixels.  The modest angular resolution might mean that a point
source triggers more than a photomultiplier, allowing for charged
cosmic ray secondary particles going through individual
photomultipliers to be eliminated.  However, a better method might
be multidish coincidence as discussed above.

The key point of this section is that area, rather than optical
quality, is the relevant dish parameter that needs to be
maximized.  Poor angular resolution can be tolerated for OSETI
because of the short pulse duration given that nanosecond time
binning is used .

In the above, we have chosen a pixel size that is small enough
that the diffuse night sky background is not much more of a
problem than the host star of a solar luminosity at 100 pc, and
this happens to be about 10$^5$ square arcseconds, the pixel size
of next generation ACT cameras.  The minimal signals we considered
for confident detection at such a distance translates to tens of
kilojoules for a nanosecond pulse of 10$^{-13}$ ster. For brighter
signals, such as the megajoule pulses considered by Horowitz and
co-workers (2001), the minimal angular resolution is a significant
fraction of a degree, and huge collecting surfaces could be made
less expensively.  The PETAL dish, serendipitously, is nearly
ideal for such purposes.

Rather than monitoring individual stars, the dish can merely look
off into space, while doing whatever else it was designed to do.
With a field of view of 12 square degrees, as per the current
design of the MAGIC camera, and an assumed range of order 100
parsec, such a dish would cover about 10$^2$ stars.

Most of these stars would be thought incapable of supporting life.
On the other hand, it is hard to know for sure what fractions to
consider and to reject. Moreover, it is hard to know {\it a
priori} what strength optical pulses are worth looking for.
Stronger pulses allow deeper searches, and the search volume could
be larger than the 10$^3$ pc$^3$ that we have envisioned here.

\section{Detecting Orbital Motion}

 Two ACT observatories on different
continents could verify that optical pulses came from a planet
orbiting the star if they simultaneously detected repeated pulses
over several months. Using the Earth as a baseline, and allowing
for the limited number of ACT's worldwide and the need for
simultaneous observing not too close to dawn or dusk, etc., one
could obtain with ns resolution an angular resolution on the order
of 10$^{-7}$ radians.  Even at a distance of 100 pc, then, the
orbital motion of an ongoing signal could be detected if the
orbital radius is 1 AU or more.  In contrast to Doppler and timing
techniques, this would not depend on the pulses being received
regularly or on the accuracy of the clocks of the ETI.

A multi-continental simultaneous observing program would increase
the collecting area, enhance the capability for coincidence-based
noise rejection (especially if there is concern about local and
regional sources of non-Poissonian noise), and would thus ease the
requirements made on each individual ACT observatory.  Thus, there
is added motivation for such a transcontinental OSETI effort.  The
discussion in this section applies as well to OSETI with smaller
area and better angular resolution.

\section{Summary and Discussion}

 We have argued that giant parabolic
dishes with modest angular resolution and large fields of view
offer an enhanced, cost effective search capability for OSETI.
Their larger collecting areas and fields of view ultimately allow
them to monitor many stars at once even while doing gamma ray
observations.  If they are dedicated to OSETI, then, unlike
smaller telescopes, they could target candidate host stars with
more sensitivity or those at greater distances.  If, as seems
probable, the number of pixels roughly equals or exceeds the
number of stars monitored at once, then an individual star thought
to be associated with a positive signal (or in any case a short
list of suspects) could be identified for further monitoring with
classical astronomical angular resolution.

The price paid for monitoring 10 or more square degrees at once is
an enormous cosmic ray background.  However, the next generation
of ACT arrays seems capable of handling this problem in principle
with suitable online software.  Moreover, an ET civilization
intent on self-broadcasting with laser pulses might design their
time profiles to stand out above this background.

\section{Acknowledgments}

We thank Profs. Sandra Faber, David Faiman, Sergey Biryukov, Noah
Brosch, Jill Tarter, Phillip Morrison, Phyllis Morrison, and
Sergey Karpov for fruitful discussions. This research was
supported by the Arnow Chair of Theoretical Astrophysics, ISF
grant 208/98-2, the Russian Foundation of Fundamental Research
(grant 01-02-17857) and the Russian Federal Program "Astronomy".

\noindent Figure caption: The PETAL optical reflecting dish in
Sede Boqer.


\begin{thebibliography}{99}

\bibitem{}Beskin, G., N. Borisov, V. Komarova, S. Mitronova, S. Neizvestny,
V. Plokhotnichenko, M. Popova (1997)  Methods and Results of an
Optical Search for Extraterrestrial Civilizations. {\it
Astrophysics and Space Science}, {\bf 252}, 51-57.


\bibitem{}Barrio, J.A., Blanchot, G., Borst, H.G., Blanch, O.,  Bosman, M.,
Bradbury, S,M.,
 Cavalli-Sforza, M.,  Chilingarian,  A., Contreras, J.L. , Cortina, J. , Dosil, M.,
Feigl, E., Ferenc, D.,   Fernandez,  E.,   Fernandez, J. Fonseca,
V., Gebauer, H.J.,Gonzalez, J.C.,  Haaf, E., Holl, I., Hrupec, D.,
Ibarra, A.,   Karle, A.,  Kornmayer,H., Krawcynski, H., Llompart,
Z., Lorenz,  E., Magnussen, N., . Mariotti, M., Martinez, M.,
Merck, M.,  Meyer, H.,   Mirzoyan, R.,   Moralejo, A.,   Moller,
H., Muller, N.,  Odeh, T.,  Ostankov, A.,  Padilla, L.,  Petry,
D., Plaga, R.,   Prosch, C.,   Raubenheimer,  C., Rauterberg,  G.,
Sawallisch, P.,   Schmidt, T.,  Turini, N., Wacker, A.,  (1998)
{\it The MAGIC Telescope design report: Report MPI-PhE/98-5.}
Max-Plank-Institut fur Physik, Munich.

\bibitem{}Horowitz, P.,  Coldwell, C.,   Latham, D.,  Papaliolios, C.,
Stefanik,  R., Wolff, J., Zajac,J. (2001).  Targeted and All-Sky
Search for Nanosecond Optical Pulses at Harvard-Smithsonian,
preprint

\bibitem{}Schwartz, R., and C. Townes 1961. Interstellar and Interplanetary
Communication by Optical Masers. {\it Nature}, {\bf 190}, 205-208.

\bibitem{}Shvartsman, V., G. Beskin, S. Mitronova, S. Neizvestny, V.
Plokhotnichenko, L. Pustilnik 1993. Results of the MANIA
Experiment: an Optical Search for Extraterrestrial Intelligence.
In {\it Third Decennial US - USSR Conference on SETI} (G.S.
Shostak, Ed.), pp. 381-390. Astronomical Society of the Pacific
Conf. Ser., {\bf 47}, San Francisco, California, USA.

\end{thebibliography}
\end{document}